\documentclass[runningheads]{llncs}
\usepackage[latin1]{inputenc}
\usepackage[TS1,T1]{fontenc}
\usepackage{lmodern,graphicx,color,url}
\graphicspath{
 {images/}
}
\usepackage{dcolumn}
\newcolumntype{d}[1]{D{+}{+}{#1}}
\usepackage{textcomp}

\begin{document}

\title{Report on the current state of the French DMLs}
\titlerunning{French DMLs}
\author{Thierry Bouche}
\authorrunning{Thierry Bouche}
\institute{Universit\'e de Grenoble~I \& CNRS,\\
Institut Fourier (UMR 5582) \& Cellule Mathdoc (UMS 5638),\\
BP 74,  38402 St-Martin-d'H\`eres Cedex, France\\
{thierry.bouche@ujf-grenoble.fr},\\ 
URL: \url{http://www-fourier.ujf-grenoble.fr/~bouche/}}

\maketitle              

\begin{abstract}
  This is a survey of the existing digital collections of French
  mathematical literature, run by non-profit organizations. This
  includes research monographs, serials, proceedings, Ph. D. theses,
  collected works, books and personal websites. 
\end{abstract}

\thispagestyle{empty}

\section{Introduction}

If we view the Digital Mathematics Library as the electronic
counterpart to the traditional mathematics' laboratory library, it
should hold a similar set of shelves: research monographs; serials;
proceedings of conferences and seminars; Ph. D. and other theses;
collected works; sometimes multimedia material (videotaped
conferences\ldots); as well as precious items that are sometimes in a
reserved area (manuscripts, correspondence, personal archives of some
great scholar\ldots). Libraries also display new acquisitions and
often stock incoming preprints. Depending on their users' profile,
they hold also a lot of non-research texts: we won't discuss these
here.

We will try to survey what has currently been done in this direction
in the small French context, by whom, and under which policies. This
survey is far from exhaustive, because of the way these efforts have
been pursued, mostly in isolation, with often very restrictive goals
and targeted user bases. If ``French content'' is to be understood
roughly as mathematical research texts mainly authored or published in
France, it can be found in many foreign projects (books in Cornell or
Michigan, journals in Project Euclid, or at Göttingen's GDZ, articles
in virtually any journal worldwide\ldots) while foreign material is
very present as well in most French projects. The French context we
focus on is to be understood in a rather vague manner, meaning that
either the digital library project or its content has strong roots in
France. 

A simple picture can be drawn as follows: Bibliothèque nationale de
France (BNF, French national library) runs Gallica~\cite{gallica}, one
of the oldest large-scale generalist digitisation programme, primarily
focused on public domain books. It has 736 books filed under
mathematics, and 5 serials. Cellule MathDoc (CMD, a mathematical
literature service unit of CNRS and University of Grenoble~I) runs
NUMDAM~\cite{numdam}, a digitisation programme primarily focusing on
mathematical journals published in France (30 journals and 29 seminar
proceedings, 3900 volumes). Thus, roughly, you'd find old books at
Gallica while you'd find recent journal articles at NUMDAM. CMD is
associated to BNF in order to synchronise their digitisation efforts,
and CMD provides an article-level catalogue of relevant Gallica
collective volumes to the extent possible within available resources,
see \cite{Ga-math}. Beside these cooperating big players, a plethora
of smaller projects exist, weakly interrelated.

\section{Books and theses}

As long as no research or higher education exceptions to copyright
laws appear, archiving and providing access to electronic versions of
recent books is a risky business. Works being typically in the public
domain 70 years after the death of the last surviving author, many
projects tend to consider that they are safe when providing only books
published up to the 19th century. More recent works have been dealt
with either on a purely commercial basis, or by smaller groups.  As a
result, a very small portion of the books published since 20th century
are archived in digital form, and curated by an independent
institution.

\subsection{Gallica}

Some 736 digitised books are present in Gallica, freely downloadable.
The content is restricted to public domain books (usually up to the
start of 20th century). But sometimes up to 70 years old ones (when
they can be considered collective works). The above number is somewhat
approximative, and has grown and decreased over time. One reason is
that, as some old books in the public domain were digitised from
recent reprints, the publisher of the reprint sometimes obtained their
withdrawal.

Some of these are written in Latin, German, Italian, many of them
being written by important foreign scientists. 

The Gallica web site \cite{gallica} has recently been updated
(nickname: Gallica~2) to a larger digital library portal (a
\emph{virtual} library), which contains references to digital
documents from other providers, some of them being smaller public
libraries while many others are only providing metadata and a preview
of their resources to Gallica users, who need to pay for full access.
Some mathematical books--mostly educational--appear that way.

\subsection{Patrimoine numérisé du Service de la documentation de
  l'université de Strasbourg}

The central library of Strasbourg university digitises some of the old
and prestigious books in its collections. This yields 139 books
covering widely mathematics from 1544 to 1917, in Latin, German,
English and French \cite{SICDStr}.

\subsection{Les bibliothèques virtuelles humanistes}

Tours university project of careful digitisation of books from the
early times of printing \cite{BVH} do contain some mathematical stuff (5 books
from 1516 to 1581, written in Latin, Italian, and French) as well as
hard sciences of that time. This project is not a small one, but its
main focus (early humanities) makes mathematics a low priority.

\subsection{PôLib}

Lille's library runs a similar project \cite{POLIB} providing us with
2 old (1634 and 1731) French mathematical books. These are typically
books that patrons of the library could not put their hands on.
Unfortunately, the image-oriented interface makes this web posting
more similar to a display cabinet exhibition, rather than a tool for
the working scholar. A complaint I could repeat for many similar
projects showcasing some library's treasures rather than helping their
dissemination and actual use.

\subsection{NUMIX}

The library of École polytechnique and friends digitised 4 books
published by that school with mostly texts of lectures given there
\cite{NUMIX}.

\subsection{Jubilothèque}

This is the scientific digital library of University Pierre \& Marie
Curie, in Paris, pursuing similar goals as the Strasbourg project
already mentioned. It doesn't have much math, but still a bunch of
theses defended at Sorbonne during the 19th century, among which the
one of Poincaré, e.g. \cite{Jubil}.

In fact, most university libraries are currently in a similar process,
setting up digital libraries underlining the local production, with a
special stress on theses, and remarkable items hosted locally.

\subsection{NUMDAM}

There are currently no books in NUMDAM, but some are planned. The
\emph{Mémoires de la SMF}, supplement to the \emph{Bulletin}, can be
viewed as a research monographs series. We intend to follow-up this
effort with reasonably copyright-free memoir series, like ``Mémorial
des Sciences Mathématiques'' (Gauthier-Villars). 

We have already digitised some 450 doctoral theses defended in France
between World Wars I \& II and will post them in a near future.

In the same vein as the useful collection of seminar proceedings, we
will also start to provide (formally) unpublished or rare monographs,
like lecture notes of summer schools that were mimeographed for
dissemination to a small audience. The difficulty here is to select
those documents that are really useful to current or future active
research: some of them are quite obvious as they have been
xerox-copied, circulated and referenced among interested circles. The
other difficulties are to contact the author for authorisation, to
find a copy that can be removed from the shelves long enough to
assemble a production batch\ldots

\subsection{Jussieu-Chevaleret mathematics' library}

There is an ongoing project to digitise about 500 out-of-print books
published by Springer group. Some of these are quite recent.  The
original goal here is not at all to contribute a new digital corpus to
the mathematicians worldwide, but to partially solve the problem faced
by two labs originally in the same campus (Jussieu, in Paris) and
sharing the same library, one of which has been relocated at some
distance (Chevaleret).

The result might be a digital collection with free access from
Jussieu and Chevaleret campuses, possibly also from the French universities.
In parallel, Springer would exploit commercially the digital files as
it does for its in-house digitised collections. This is an example
where the economical balance between public funding, public interest,
and publisher's partial ownership over the collection results not in
chronological frontiers (moving walls\ldots), but geographical ones.

\section{Journals and other serials}

More than 20 research journals in mathematics are currently published
in France.  Most, but not all of them have plans for independent
archiving by a library service.  Copyright policies seem to be less
constrained for journal articles than for books, as the main value of
journals to publishers comes from keeping the number of subscribers
high, rather than collecting article-level pay-per-view, and authors
are eager to get as much dissemination of their research findings as
possible.

\subsection{NUMDAM}

The current online ``French'' content consists of 27~serials,
28~seminar series ; summing up 30\,000~articles spanning over more
than 600\,000~pages; more than 10 serials are planned for the
immediate future.

Almost all recent French periodicals are included, many seminars.
Subject area coverage spans from theoretical physics to computer
sciences, optimisation, statistics, applied mathematics and the core
of pure mathematics. 

The NUMDAM website hosts the output of the NUMDAM digitisation
programme, as well as born-digital articles supplied by their
publishers.  New content is acquired for 13 journals and 3 seminars
from 4 publishing platforms (Elsevier, Springer, CEDRAM, EDP
Sciences--the list should include project Euclid and SMF in a near
future, as two journals previously published by Elsevier have moved).

Very recent articles are present in the library, which acts as a
portal (indexing, discovery service) to the publisher's platforms for
them. The full text is available as DjVu and PDF, and freely
accessible locally after a moving wall, which lasts typically 5 years.

\subsection{Gallica}

Gallica content is still growing, with a bias toward venerable and
generalist journals, such as \emph{Journal des Sçavans}, \emph{Journal
  de l'École polytechnique} or \emph{Répertoire bibliographique des
  sciences mathématiques}.  But it hosts also journals published by
Elsevier since 1997, when Gauthier-Villars' stock was acquired, with a
70 years ``moving wall'' like \emph{Bulletin des sciences
  mathématiques} (Darboux), \emph{Journal de mathématiques pures et
  appliquées} (Liouville). All series of \emph{Comptes rendus de
  l'académie des sciences} are also available (not yet exhaustively)
up to 1993.

The reason for this 70 year delay is the following: Gallica has the
policy that, if a  journal volume is digitised as a whole, and
metadata is collected at the volume level only, then it can be
considered as a collective work, for which copyright expires 70 years
after publication, irrelevant of authors' destinies.

\subsection{Orsay library}

Orsay library scanned the entire collection of the \emph{Publications
  mathématiques d'Orsay} whose content type is comparable to
Springer's Lecture Notes in mathematics. It is planned to be added to
NUMDAM eventually. Meanwhile, a dedicated browsing interface has been
set up by MathDoc~\cite{PMO}.

Many laboratories in France had such kinds of publication series (it
is indeed a prefiguration of institutional repositories: they may
contain preprints or preliminary versions of books that have since
been published, lecture notes, proceedings of some sort, and original
works unpublished elsewhere, like works of Ecalle on resurgent
functions).  Few have been digitised.

\subsection{EDP Science/HAL}

The French publisher EDP Sciences (which derives from the publishing
department of the French society of Physics) has digitised on its own
the full backrun of \emph{Journal de Physique}, \emph{Radium}, and
related series. This ends up as a separate package \cite{jdphys}
marketed under a similar scheme as Springer Online Journal Archives or
Elsevier Backfiles. But this could also introduce the concept of green
archive, as the whole content (with strictly minimal metadata and no
navigation devices) has been copied into the French open archive HAL
as a separate collection \cite{jdphyshal}.

\subsection{INRIA Rocquencourt}

Revue \emph{Modulad}, a journal of statistics dedicated to data analysis is
published by INRIA at Rocquencourt. It is open access, and all early
volumes have been digitised and posted as well together with the
current edition of the journal \cite{modulad}.

\smallskip

Journal digitisation being more labour intensive than books
(especially article-level metadata generation), there are not so many
small-scale digitisation run by individuals or small voluntary groups.
Nevertheless, at least another data analysis journal has been
partially scanned and posted in a relatively amateur fashion:
\emph{Les cahiers de l'analyse des données} \cite{CAD}.

\subsection{Foreign projects}

Some French serials and books are available from foreign projects. The
institutions have not always been contacted about these\ldots{} For
instance, the \emph{Bulletin astronomique}, published by Paris'
observatory since 1870, has been scanned and posted (as a raw list of
images) by the SAO/NASA Astrophysics Data System at Harvard
\cite{ADS}.

Göttingen digitisation centre has digitised the seminar proceedings,
an ancestor of the current \emph{Journal de théorie des nombres de
  Bordeaux} \cite{STNB}.

Some live journals published by French institutions are published
abroad. For instance, \emph{Annales de l'institut Henri-Poincaré,
  Probabilités et Statistiques} are now published by IMS and
electronically delivered by project Euclid \cite{AIHPB}; the very
recent \emph{Journal of the Institute of Mathematics of Jussieu} is
published in United-Kingdom by Cambridge University Press
\cite{JIMJ}.

\subsection{Unarchived or unarchivable}

Some live journals still have no plans for an independent safe
archive, curated independently for the public good. Some are probably
too recent or not yet enough established as strong references to care,
but this might be an issue if they realise it was necessary after the
first failure to locate a needed resource. This is the case, for
instance, of the \emph{International Journal On Finite Volumes}, a
small independent journal currently hosted on a server in Marseille
\cite{IJFV}; or of \emph{Journal of the Institute of Mathematics of
  Jussieu} \cite{JIMJ}.

Worse is the case of journals published by Hermès/Lavoisier, like
\emph{European Journal of Control} \cite{EJC}. The articles are posted
in some sort of PDF format, protected by digital rights management (DRM)
techniques, so that they can't be read out of a very strict
environment (Adobe reader software, HTTP connectivity, etc.). We hope
that the content of these journals' articles won't be useful when the
current document and network technology will have been lost!

\section{Electronic collected works}

Collecting seminal works from influential scholars has always been a
flourishing activity, and often dedicated areas in libraries are
reserved to store them. Variations on this theme have been one of the
most prominent activity since the early scholarly Internet (even
before the World Wide Web was invented and deployed): historians
collect resources on a period, on some scientist; researchers collect
articles on their subject; etc. These collections belong to a grey
area between informal publishing and personal libraries.

\subsection{NUMDAM/SMF/Polytechnique}

The Société mathématique de France, in cooperation with École
polytechnique, intended to publish a volume comprising selected works
of Laurent Schwartz. As NUMDAM had already 117~articles available, we
digitised 61 more so that a more complete CD-ROM can be bundled to the
paper volume. This should be eventually integrated to the NUMDAM
website in a special area, where other collected works are expected,
such as those of Charles Ehresmann.

\subsection{Grothendieck circle}

As an example of independent collective initiative, let us mention a
project of electronic collected works of Alexander Grothendieck by a
group of mathematicians calling themselves the Bourbakistas. They
collect digital articles from existing libraries such as JSTOR,
NUMDAM, etc. When important resources are not already available (such
as the Séminaires de géométrie algébrique du Bois-Marie), they
digitised them and served them on the Web.

This is the sort of library you organise in your own office and
possibly open to colleagues.

\subsection{Personal collected works}

In 2001, the IMU Executive Committee endorsed a call to all
mathematicians to make available electronically as much of their own
work as feasible \cite{IMUPCW}. This is known as the  suggestion that
each mathematician should edit himself his own ``personal collected
works'' as a dedicated web page with downloadable items.

I don't know whether this call, vanity, or the necessity of
self-promotion in our competitive world explains the large number of
works mathematicians upload on their home pages, but it makes probably
already the biggest virtual collection of digital mathematics.

However, mathematicians do not take care of details in the same way as
librarians, or publishers: it is not obvious how much these
collections should be trusted, and whether we can rely upon them for
the long term archiving and access to the literature. For instance, a
very well organised such collection is provided by Alain Connes on his
personal website \cite{ACPCW}. But virtually any mathematician's
home page presents a variation on this theme. 

You can find there almost every paper he published since 1998, as well
as two books (the seminal \emph{Noncommutative Geometry}, as published
by Academic press in 1994, and an ongoing work-in-progress). When you
look at the actual PDFs of the papers, you are faced with the fact
that some are scanned from papers, some are arXiv preprints, some are
publisher's final proofs, some are publisher's final printouts, some
are renamed files copied from digital libraries. This serves the basic
service expected by the author to help people visiting his home page
dig into his work. Possibly, coupled with Google Scholar or similar,
this provides some ``green open access'' to his \oe uvre, and adds to
its (already vast) visibility.

\section{Archives and rare items}

Some manuscripts have been digitised, like Laurent Schwartz' personal
archives at Polytechnique, Nicolas Bourbaki's by a defunct CNRS unit,
now curated by MathDoc \cite{BBKI}.

Another example is the project to edit unpublished (or even unfinished)
works by Michel Herman after his death by Jean-Christophe Yoccoz at
Collège de France \cite{JCYMH}. 

Also, multimedia is emerging: one can find on the Web collections of
videotaped conferences, mathematical movies, animations, javascripts
or other applets. I am not aware of many systematic attempts to archive
these new media into an organised and well maintained library.
Electronic Geometry Models is an interesting one \cite{EGM}, which was
indexed recently in Zentralblatt.

\section{Conclusions}

We tried to give an overview of the existing archives of the French
mathematical literature, which are run by not-for-profit organisations
or individuals.  The ``French'' area is a small part of the whole
mathematical corpus (20 core journals among 600, no more such
important publishers as Gauthier-Villars). Yet the number of
formalised projects, and myriad of grey ones, might give an idea of
the complexity of the task of registering the ongoing efforts and
available items. Not to mention the question of settling whether
apparent duplicates are identical, different versions, or bear
different mathematical meaning; whether a given copy is legal or not,
mathematically validated or not, and by whom.

Of course, a larger body of texts is available through commercial
offers, most of them being run by publishers through a variety of
business models. The notable exception being the Google Books project
\cite{Ggbk}, which is still in beta stage, so that it is probably too
early to derive strong conclusions on its quality from the way it
currently operates. Google Books has virtually digitised all the
mathematical content considered in the above lines, possibly many
times. For instance, 114 volumes of the \emph{Journal de mathématiques
  pures et appliquées} have been digitised, some of them multiple
times at different locations.  The copyright of many, but not all, of
them has been attributed to its current publisher (Elsevier), who runs
it only since 1997. We don't know whether this explains the fact that
many of them, now in the public domain, such as the 1865 volume,
cannot be accessed fully.

In the digital realm, the divide between the traditional library and
publishing functions tends to fade away. Some of the collections we
mentioned can equally be considered as personal libraries or \emph{de
  facto} Web editions. An important divide exists nevertheless: a
publisher creates new content, brands it and market it while a library
acquires new content in order to archive it and preserve it over the
long term, and manage long term access to its patrons.  While a number
of business models exist for publishers, libraries cannot be run for
the sake of profit. Publishers look forward and complete most of their
commitments when a new item is output, while libraries start at this
point. Digitisation projects are more similar to publishers in this
respect, as they create a new product (a digital version of items
previously only available on paper) which can be marketed \emph{per
  se}. In fact most independent web posting activities should rather
be seen as ``grey'' publishing. In France, some kind of ``commercial
digital library'' ventures have emerged. An example is given by the
NUMILOG society \cite{numilog}, which contracts with university
libraries so that their patrons have a certain global number of
authorised downloads, with a time limit enforced through DRMs.
Bourbaki books are available in this offer. In this brave new model,
the university library is only a mediator between the society, driven
by profit, which is where the master unrestricted files are hosted,
and its patrons.

If we want a safe archive of the mathematical corpus, some sort of
traditional library has to be set up. We will have to find a path
between the all-inclusive view that any digital mathematical text
referenced in, e.g., Scirus or Google Scholar forms part of a huge
virtual library of mathematics, and the too-picky view that a small
club of large memory institutions can handle physically more than a
small percentage of the whole corpus.

\end{document}